\documentclass [12pt]{article}  
\usepackage{epsfig}
\begin{document}
\begin{titlepage}
\def\today{\number\day
           \space\ifcase\month\or
             January\or February\or March\or April\or May\or June\or
             July\or August\or September\or October\or November\or December\fi
           \space\number\year}
\def\thisday{15 May 1997}
\hbox{
\vtop{
\hbox to \hsize{\hfill Fermilab-Conf-97/141-T}
\hbox to \hsize{\hfill hep-ph/9705384}
\hbox to \hsize{\hfill \thisday}}}

\vspace{1.5cm}
\begin{center}

{\Large 
Next to Leading Order Three Jet Production at Hadron Colliders\large
\footnote{Talk presented at XXXIInd Rencontres de Moriond, QCD and
High Energy Hadronic Interactions, Les Arcs, Savoie, France, 22 - 29
March, 1997.}
\\[1.0cm]} 
{ \sc William B. Kilgore \footnote{\tt kilgore@fnal.gov}\\[2.mm]}
{ \it Fermi National Accelerator Laboratory\\ P.O. Box 500
\\Batavia, IL 60510, USA}

\end{center}
\vspace{1.5cm}
\begin{abstract}
\noindent
I present results from a next-to-leading order event generator of
purely gluonic jet production.  This calculation is the first step in
the construction of a full next-to-leading order calculation 
of three jet production at hadron colliders.  Several jet algorithms
commonly used in experiments are implemented and their numerical
stability is investigated.  A numerical instability is found in the
iterative cone algorithm which makes it inappropriate for use in
fixed order calculations beyond leading order.
\end{abstract}
\vfill
\end{titlepage}

\section{Introduction}
\hskip\parindent
In this talk I report the first step in constructing a 
Next-to-Leading Order (NLO) three jet event generator for hadron
colliders \cite{KG}, work that has been done in collaboration with
Walter Giele.  This first step involves the construction of the purely
gluonic contribution to this cross section.  It is expected that this
calculation will have a number of phenomenological applications.
In this talk, I will present results on the numerical stability of
various jet algorithms and on their applicability to the comparison of
theory to experiment.  In particular, I will find that the iterative
cone algorithms currently used by CDF and D0 have an infrared
instability which precludes their use in fixed order calculations
beyond leading order.  In section 2, I will describe the methods and
techniques used in the event generator, in section 3, I will describe
the jet algorithms used, in section 4, I will present the results of
our study and in section 5, I will summarize our findings.
\section{The method}
\hskip\parindent
The calculation of gluonic three jet production at next-to-leading
order combines the $gg\rightarrow ggg$ process computed to one loop
\cite{BDKa} with the $gg\rightarrow gggg$ process computed at Born
approximation \cite{PT,Ku,GuKa,BGa,MPX,MP}.  individually, each of
these contributions contains infrared singularities.  Only the sum of
the two contributions is finite and therefore only the sum provides a
meaningful calculation of the cross section.  The one loop amplitudes
are infrared singular because of massless partons going ``on shell''
within the loops.  These singularities appear as double and single
poles in the dimensional regulator $\varepsilon$ multiplying the Born
amplitude.  The bremsstrahlung term ($gg\rightarrow gggg$) becomes
singular when one final state parton becomes very soft or when one
parton becomes highly collinear with another.  In either case, the
singularities appear when there is a loss of parton resolvability,
that is when the four parton final state is indistinguishable from a
three parton final state.

The cross section for any number of {\it resolved \/} partons is
finite and well defined at each order in perturbation theory.  All
that is needed is a resolution criterion.  This resolution criterion
can take many forms, from a simple invariant mass cut to a full blown
fragmentation function.  For this study a simple invariant mass
resolution criterion, $s_{\rm min}$, suffices \cite{GG,GGK}.  For two
partons $i$ and $j$, if $s_{ij} > s_{\rm min}$, the two partons are
said to be resolved from one another, while if $s_{ij} < s_{\rm min}$,
the partons are said to be unresolvable and the event falls into the
soft or collinear region of phase space.  Note that for massless
partons $s_{ij} = 2E_iE_j(1-\cos\theta_{ij})$, so that the $s_{\rm
min}$ criterion regulates both soft ($E_i\rightarrow 0$) and collinear
($\theta_{ij}\rightarrow0$) configurations simultaneously.

If $s_{\rm min}$ is chosen to be sufficiently small, the concept of
parton resolution should not interfere with that of jet resolution.
The identification of a parton as ``resolved'' does not preclude it
from being clustered with other partons by the jet algorithm.  It
simply organizes the partonic calculation and allows for the
cancellation of the infrared singularities.  This cancellation is
accomplished by an improved version the phase space slicing method
\cite{Owens}, in which unresolvable configurations are removed from
the bremsstrahlung calculation, the true matrix elements and
unresolved phase space are replaced by their asymptotic infrared forms,
and are integrated analytically, yielding $2\rightarrow3$ parton
configurations with exactly the right singularities needed to cancel
those of the one loop terms as well as a residual logarithmic
dependence on $s_{\rm min}$.  When these terms are combined with the
one loop contribution, the result is finite but explicitly $s_{\rm
min}$ dependent.  The resulting $2\rightarrow4$ contribution is also
finite but is implicitly $s_{\rm min}$ dependent, since an $s_{\rm
min}$ dependent volume of phase space has been sliced out or it.  This
$s_{\rm min}$ dependence cancels when the $2\rightarrow3$ and
$2\rightarrow4$ contributions are combined, providing an important
cross check that the rearrangement of terms has been handled
properly.

The improvement on the slicing method introduced in this calculation,
involves correcting for the approximation of using the infrared
asymptotic forms of the matrix elements and phase space.  We refer to
this improved slicing algorithm as the subtraction method.  If $s_{\rm
min}$ is chosen to be very small, the asymptotic approximations will
be very good, and the subtraction method will not offer a significant
improvement over slicing. However, since each term depends
quadratically on the logarithm of $s_{\rm min}$, the magnitude of the
each term, and therefore the magnitude of the cancellation between the
two terms grows as $s_{\rm min}$ becomes small.  A large cancellation
requires that each term be computed to very high precision which
demands a heavy toll in computer resources.  For practical reasons,
one would therefore like to use the largest possible value of $s_{\rm
min}$.  It is expected that the subtraction method will allow us to
use a larger value of $s_{\rm min}$ than is possible with the slicing
method.

\section{Jet Algorithms}
\hskip\parindent
The purpose of the jet algorithm is to quantify certain topological
features of hadronic energy flow in scattering processes.  By
identifying high transverse momentum hadronic clusters in collisions
we can make a connection with the underlying partonic scattering and
apply perturbative QCD to predict the cross section.  The NLO three
jet calculation provides new sensitivity to the details of jet
algorithms.  Leading order calculations are not sensitive to the
details of jet algorithms, they simply use the algorithms to 
identify hard scattering processes and eliminate configurations in
which partons are clustered together.  With minimal tuning, all jet
algorithms can be made to give the same results.  The only previous
NLO jet calculation for hadron colliders is for two jet production
\cite{EKSjet,GGK}.  Since the relevant final states contain at most
three partons, kinematics forbids the clustering of more than a single
pair, so that this calculation places no greater demand on the jet
algorithm than leading order calculations.  In the NLO three jet
calculation, there are as many as four partons in the final state, and
it is possible to obtain double clusterings, resulting in only two
jets in the final state. Such configurations belong to the NNLO two
jet cross section and are excluded from the NLO three jet calculation.
It is at the multiple clustering boundary that the differences in jet
algorithms become apparent.

In this study, we consider four common jet clustering algorithms:
\begin{itemize}
\parskip=0pt
\item[(a)] The ``fixed-cone'' algorithm used by UA2 \cite{UA2alg}.
\item[(b)] The ``iterative-cone'' algorithm, used by CDF \cite{CDFalg} 
           and D0 \cite{D0alg}.  
\item[(c)] The ``EKS'' algorithm, used in NLO 1-jet and 2-jet 
           inclusive calculations \cite{EKSRsep}.
\item[(d)] The ``$K_T$'' algorithm \cite{KtalgTh}, 
           under study by CDF and D0 \cite{KtalgExp}.  
\end{itemize}
The fixed-cone algorithm is phenomenologically inspired; one simply
draws a cone of radius $R$ (in pseudorapidity $\eta$ -- azimuthal
angle $\phi$ space) around the highest transverse energy ($E_T$)
cluster and all hadronic energy within that cone is assigned to the
jet.  The final jet coordinates are determined by the $E_T$ weighted
center of the jet.  The iterative cone algorithm seeks to find the
best possible cone by starting as before but once the $E_T$ weighted
center is found, a new cone is drawn around that center and a new
$E_T$ weighted center is found.  This process is iterated until a
stable cone axis is found.  The EKS algorithm is a modification of the
``Snowmass Accord'' \cite{Snowmass}, in which one computes the
hypothetical $E_T$ weighted axis between two partons, and if both
partons are within radius $R$ of the axis, they are clustered into a
jet unless the partons are more than $R\times R_{\rm sep}$ ($R_{\rm
sep} \le 2$) apart.  This simulates the iterative-cone algorithm by
assuming that it always find the optimum jet-axis.  The $K_T$
algorithm is inspired by the parton shower dynamics of QCD.  Jets are
built up by merging those clusters which are ``closest'' to one
another in $d_{ij} = \min\{E_{Ti}, E_{Tj}\}\Delta R_{ij}$.  For a
complete description of these algorithms and our implementation, see
the references cited above as well as reference \cite{KG}.

The numerical stability of the four jet algorithms is related to the
degree to which the algorithm is sensitive to soft radiation, or in
other words the infrared stability of the particular algorithm.  For
the method of resolved partons, as is used in this paper, infrared
stability is related to the extent to which the results are
independent of the the resolution parameter $s_{min}$.  This
dependence is shown in fig.~\ref{fig:smin} and will be discussed in
the next section.

\section{Numerical Results}
\hskip\parindent
For the numerical results in this section we used the CTEQ3M
\cite{CTEQ3M} parton distribution functions (PDF's), fixed
renormalization and factorization scales of 100 GeV and a center of
mass energy of the $p\bar{p}$-system equal to 1800 GeV.  To select
events we required one jet with $E_T > 50$ GeV and at least two other
jets with $E_T > 20$ GeV, all in the rapidity region, $|\eta| < 4$.

\begin{figure}[t]
\hbox{\epsfxsize=225pt\epsfbox{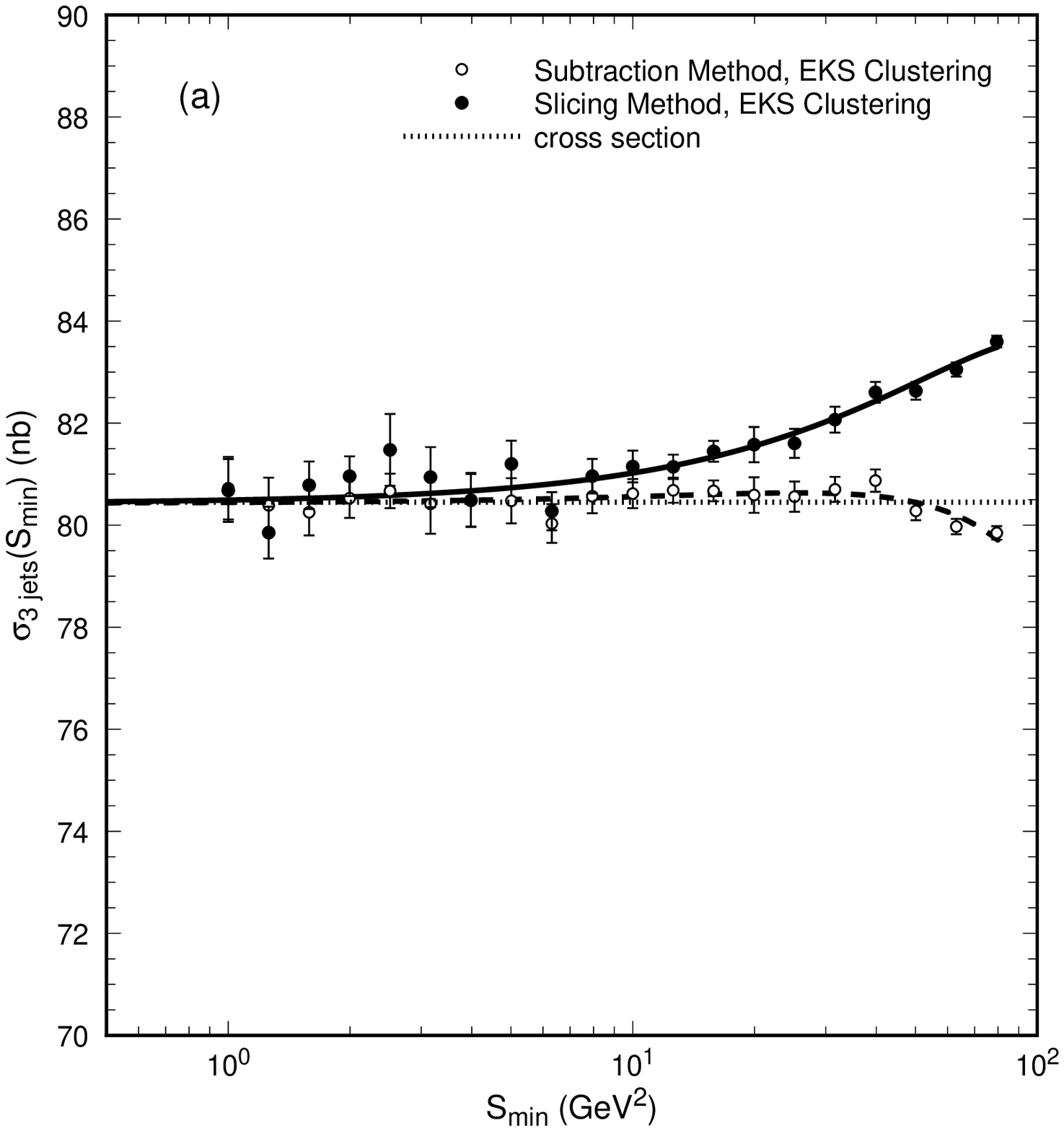}
      \epsfxsize=225pt\epsfbox{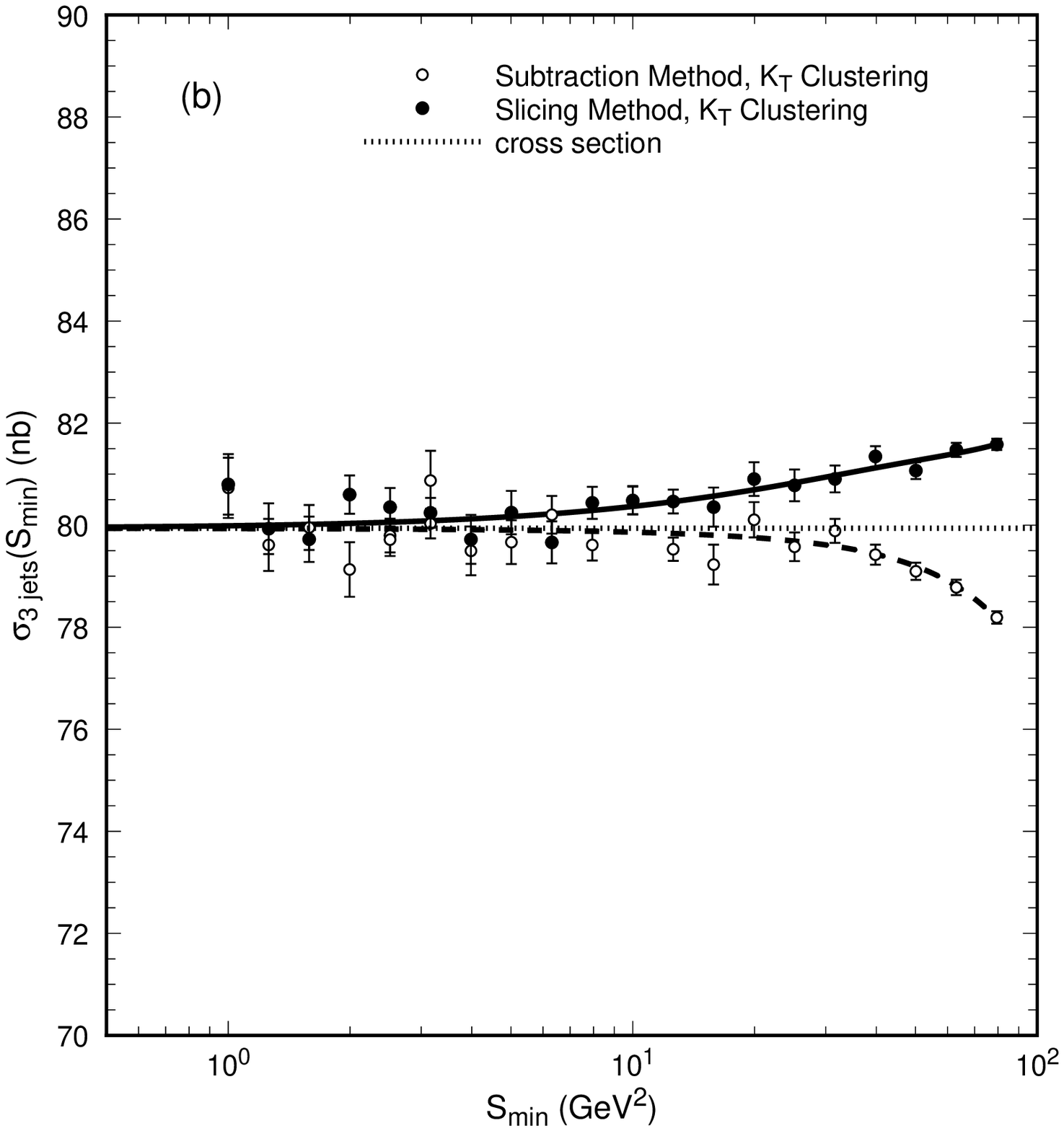}}
\vspace{25pt}
\hbox{\epsfxsize=225pt\epsfbox{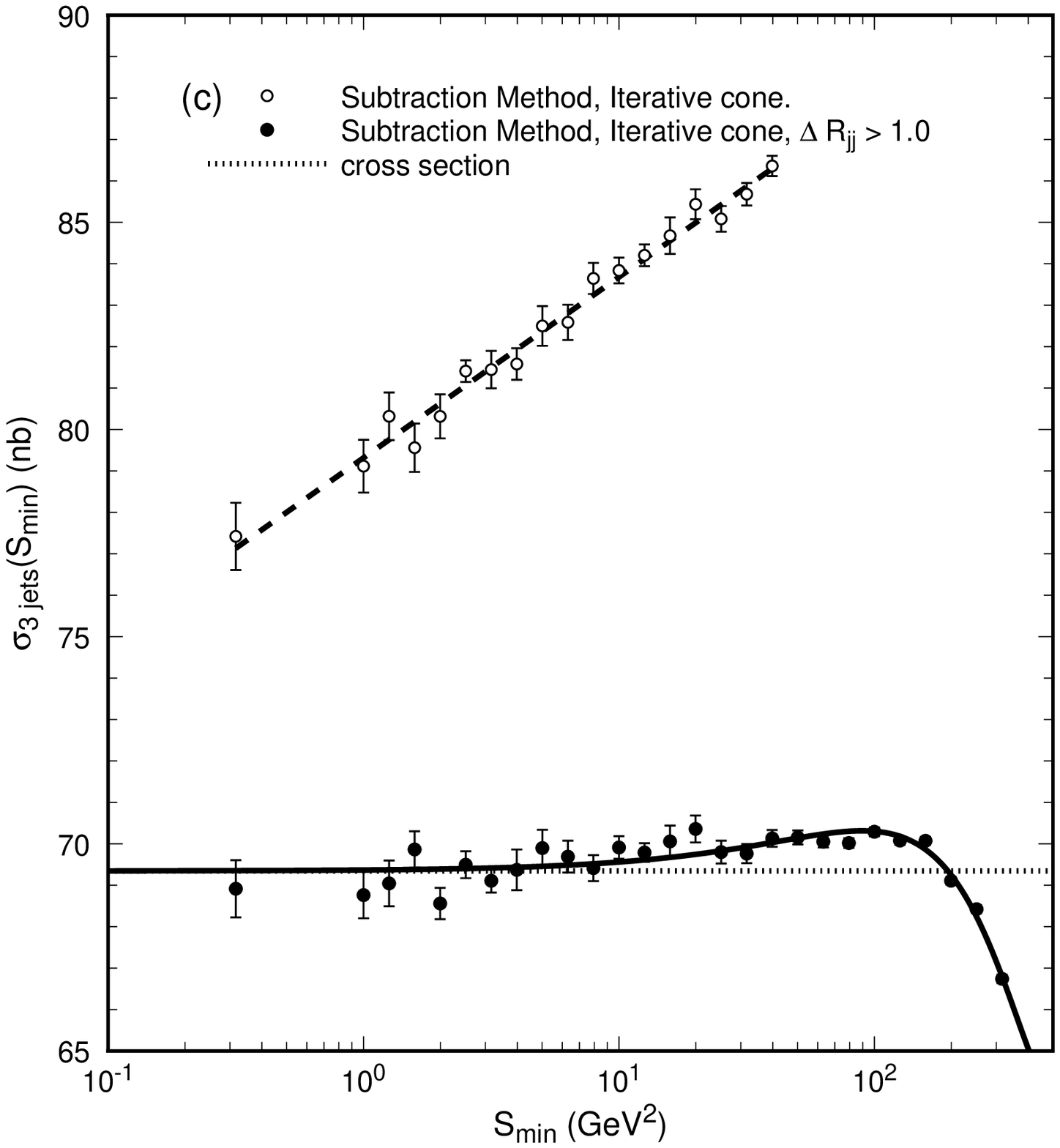}
      \epsfxsize=225pt\epsfbox{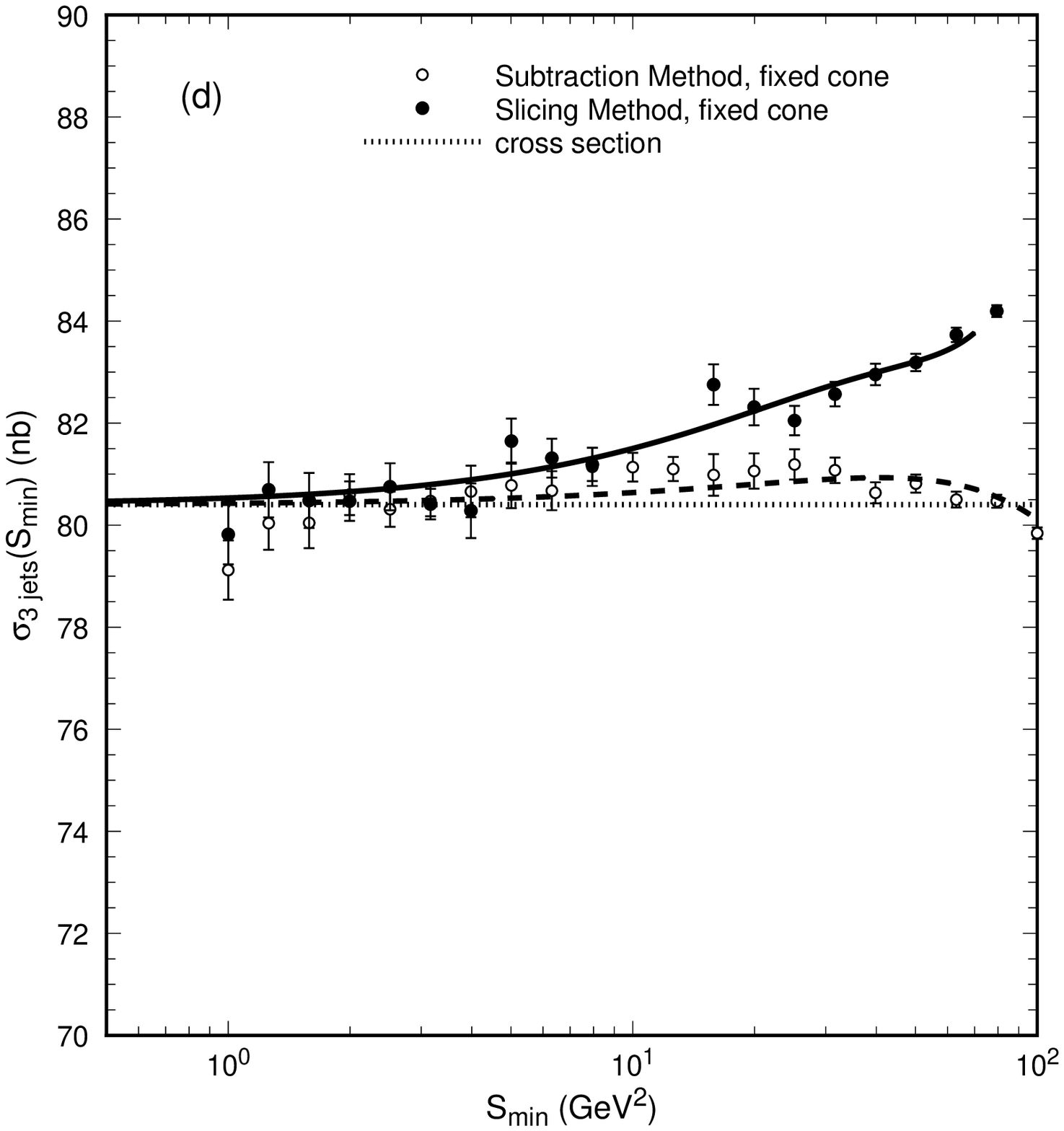}}
\caption{$\sigma_{\rm 3 jets}$ vs. $s_{min}$ for different jet
algorithms and numerical methods.}
\label{fig:smin}
\end{figure}

The first consideration is the $s_{min}$-dependence of the cross
section and the determination of the range of $s_{min}$ for which the
approximations made in the different numerical methods are valid.  The
results are shown in fig.~\ref{fig:smin} for both the slicing and
subtraction methods and all four jet algorithms.  The fixed cone, EKS
and $K_T$ algorithms behave as expected and it is clear how to
choose $s_{min}$ for them.  For the slicing method one must choose
$s_{min}$ smaller than 1 GeV$^2$ in order to get the correct answer.
As expected the subtraction method allows us to choose larger values
of $s_{min}$, though the value should still not be larger than 10
GeV$^2$.

We now consider the iterative cone algorithm. As can be seen in the
upper curve of
fig.~\ref{fig:smin}c, the cross section displays a clear logarithmic
dependence on $s_{min}$.  This means that the cancellation is failing
and that the algorithm is not infrared safe in that we can change the
jet multiplicity by adding a soft parton somewhere in the event.  This
behavior can be understood in terms of QCD dynamics and the iterative
nature of the algorithm.  QCD prefers, kinematics permitting, to
radiate in a collinear fashion.  The preferred three jet
configuration, therefore, is to have a single hard jet balanced by two
narrowly separated jets, rather than to have three equally separated
jets. Thus, for any cone size $R$, the most preferred configuration is
to have two jets separated by $R+\epsilon$.  For the tree level and
virtual contributions this is a three-jet event.  The situation should
not change if we create a four parton configuration by adding a soft
parton in between the two adjacent hard partons (as preferred by QCD's
propensity for collinear radiation), and in fact it does not change
for any of the jet algorithms other than the iterative cone. The soft
parton gets clustered with one of the hard partons, slightly changing
the jet parameters, but not affecting the jet multiplicity.

In the case of the iterative cone however, one of the two hard partons
will cluster with the soft parton thereby shifting its jet-axis to
within R of the other parton.  Because the algorithm is iterative, the
two clusters will subsequently be merged further into a single jet
resulting in a two jet final state.  Thus, we have changed the jet
multiplicity by adding an arbitrarily soft parton to the event and as
a result, the algorithm is infrared unstable and cannot be used in
higher fixed order perturbative QCD.  This change in jet multiplicity
comes about because this is a fixed order calculation involving a
delicate balance of separately divergent pieces.  Real jets in an
experiment are sprays of hadronic energy, not single partons.  Thus,
soft radiation between jets would not cause all of the energy to
become concentrated into a single jet.  More likely, it would simply
increase the overlap of the two jets and therefore slightly shift
their energy distributions.  Thus, the algorithm may appear to be
experimentally stable, but it is nonetheless clear that we cannot use
it within the NLO calculation.

Note that this result does not make the one- and two-jet inclusive
cross sections infrared unstable since in those cases we do not have to
resolve three-jet configurations.  Both CDF and D0 have compared their
multi-jet data (i.e. more than two jets in the final state) with LO
Monte Carlos \cite{D0alg,CDF2}.  It is interesting to note that the
experiments have in fact added an additional cut to their multi-jet
cross section in order to make these comparisons. This cut requires
all the jets in the event to be further apart then their cone-size of
$R=0.7$. For CDF this cut was $\Delta R_{jj} > 1.0$, while for D0 the
requirement is $\Delta R_{jj} > 1.4$. This additional requirement in
the jet algorithm changes the $s_{min}$-dependence of the cross
section dramatically, as can be seen clearly in the lower curve of
fig.~\ref{fig:smin}c. In fact the behavior is now very similar to the
other three algorithms. This is no surprise since with this additional
selection cut the infrared instability is removed. This means that the
iterative cone algorithm needs to be augmented with a jet separation
cut in order to be an infrared safe jet algorithm.


\section{Conclusions}
\hskip\parindent
In this talk I have presented results on the purely gluonic
contribution to the NLO 3-jet cross section.  All of the techniques
used can be readily applied to the quark contributions.  I have shown
that the subtraction method significantly improves the numerical
performance of the phase space slicing method.

All of the relevant experimental jet algorithms were implemented in
the NLO 3-jet event generator and their radiative effects studied. For
the iterative cone algorithm it was necessary to augment the algorithm
with an additional jet separation cut in order to obtain infrared
stability. Both CDF and D0 already apply such a cut in their multijet
analysis, though the reason is the inefficiency of the
cluster algorithm instead of the theoretically motivated removal of
the infrared instability.  The other jet algorithms behaved properly
and no additional cuts were needed.

\setcounter{secnumdepth}{0} 
\vskip 0.4cm plus 0.1cm minus 0.1cm\noindent
%
Fermilab is operated by Universities Research Association, Inc., under
contract DE-AC02-76CH03000 with the U.S. Department of Energy.
%

\vfil\eject

\end{document}